\documentclass[twocolumn, twocolappendix]{aastex63}
\usepackage[italicdiff]{physics} 
\usepackage{tcolorbox}
\usepackage[utf8]{inputenc}
\usepackage{amssymb,natbib,graphicx,color,ctable, amsmath,array,enumitem,kantlipsum}
\usepackage{scalerel}
\usepackage{hyperref}
\hypersetup{
    colorlinks=true,
    linkcolor=blue,
    citecolor=blue,
    filecolor=blue,
    urlcolor=blue
}

\usepackage[encapsulated]{CJK}
\usepackage{ucs}
\newcommand{\cntext}[1]{\begin{CJK}{UTF8}{bsmi}#1\end{CJK}}

\newcommand\bs[1]{\boldsymbol{#1}}
\newcommand\hi{H\protect\scaleto{\,$I$}{1.2ex}}

\newcommand\hiv{H\protect\scaleto{$I$}{1.2ex}}

\newcommand\ehiv{\text{\hiv}}
\newcommand\rmol{$R_{\rm mol}$}
\renewcommand{\textbf}[1]{#1}

\usepackage[flushleft]{threeparttable}
\usepackage{lineno}
\begin{document}

\defcitealias{HSvD21}{HSvD21}
\defcitealias{Gurman2025}{G25}
\defcitealias{Ostriker2022}{OK22}

\title{The GHOSDT Simulations: II. Missing H$_2$ in Simulations of a Self-Regulated Interstellar Medium}
\author[0009-0004-2434-8682]{Alon Gurman}
\affiliation{School of Physics \& Astronomy, Tel Aviv University, Ramat Aviv 69978, Israel}
\author[0000-0002-9235-3529]{Chia-Yu Hu (\cntext{胡家瑜})}
\affiliation{Department of Astronomy, University of Florida, 211 Bryant Space Sciences Center, Gainesville, FL 32611 USA}
\affiliation
{Institute of Astrophysics and Department of Physics, National Taiwan University, No. 1, Sec. 4, Roosevelt Rd., Taipei 10617, Taiwan}
\author[0000-0002-1655-5604]{Michael Y. Grudi\'{c}}
\affiliation{Center for Computational Astrophysics, Flatiron Institute, 162 5th Ave, New York, NY 10010, USA}
\author[0000-0001-8867-5026]{Ulrich P. Steinwandel}
\affiliation{Center for Computational Astrophysics, Flatiron Institute, 162 5th Ave, New York, NY 10010, USA}
\author[0000-0001-5065-9530]{Amiel Sternberg}
\affiliation{School of Physics \& Astronomy, Tel Aviv University, Ramat Aviv 69978, Israel}
\affiliation{Center for Computational Astrophysics, Flatiron Institute, 162 5th Ave, New York, NY 10010, USA}
\affiliation{Max-Planck-Institut f\"{u}r Extraterrestrische Physik, Giessenbachstrasse 1, D-85748 Garching, Germany}
\correspondingauthor{Alon Gurman}
\email{alongurman@gmail.com}

\begin{abstract}
Observations in the Galaxy and nearby spirals have established that the \hi{}-to-H$_2$ transition at solar metallicity occurs at gas weight of $P_{\rm DE}/k_B\approx 10^4 \ \rm K \ cm ^{-3}$, similar to solar neighbourhood conditions. Even so, state-of-the-art models of a self-regulated interstellar medium underproduce the molecular fraction ($R_{\rm mol}\equiv M_{\rm H_2}/M_{\ehiv{}}$) at solar neighbourhood conditions by a factor of $\approx2-4$. We use the GHOSDT suite of simulations at a mass resolution range of $100-0.25\ M_{\odot}$ (effective spatial resolution range of $\sim 20-0.05\ \rm pc$) run for 500 Myr to show how this problem is affected by modeling choices such as the inclusion of photoionizing radiation, assumed supernova energy, numerical resolution, inclusion of magnetic fields, and including a model for sub-grid clumping. We find that \rmol{} is not converged even at a resolution of 1 $M_{\odot}$, with \rmol{} increasing by a factor of 2 when resolution is improved from 10 to $1\ M_{\odot}$. Models excluding either photoionization or magnetic fields result in a factor 2 reduction in \rmol{}. The only model that agrees with the observed value of \rmol{} includes our sub-grid clumping model, which enhances \rmol{} by a factor of $\sim3$ compared with our fiducial model. This increases the time-averaged \rmol{} to $0.25$, in agreement with the Solar circle value, and closer to the observed median value of $0.42$ in regions comparable to the solar neighbourhood in nearby spirals. Our findings show that small-scale clumping in the ISM plays a significant role in H$_2$ formation even in high-resolution numerical simulations.
\end{abstract}


\section{Introduction}
Stars form in dense, cold, self-gravitating gas clouds \citep{McKee2007,Leroy2008,Tacconi2020,Chevance2023}. As it cools and condenses, the star-forming interstellar medium (ISM) converts an increasing fraction of its hydrogen gas mass into molecular H$_2$ form. This conversion, dubbed the \hi{}-to-H$_2$ transition, is enabled by the attenuation of stellar far-ultraviolet (far-UV) radiation via dust and H$_2$ self-shielding \citep{Hollenbach1971,Black1987,LePetit2006,Krumholz2008,Sternberg2014}. In the Galaxy, the \hi{}-to-H$_2$ transition has been observed through absorption line spectroscopy \citep{Carruthers1970, Savage1977,Gillmon2006a,Rachford2009} in diffuse gas and 21 cm and CO rotational line emission in dense gas \citep{Wilson1970,Lee2012,Lee2015}, showing a steep rise in the H$_2$ gas surface density at visual extinctions $A_{\rm V} \sim 0.5$ (with some variation) corresponding to gas surface densities $\sim 10\ M_{\odot}\ \rm pc^{-2}$ for solar metallicity. \citet{Nakanishi2003,Nakanishi2006,Nakanishi2016,Yoshiaki2016} mapped out the radial distribution of the H$_2$ and \hi{} surface densities in the Galaxy. At the Solar circle (galactocentric distance $\approx 8\ \rm kpc$), they found 
that \rmol{}, the ratio of H$_2$ to \hi{} mass, is in the range 0.19--0.26, increasing steeply inwards and towards the galactic central molecular zone. Resolved studies of \hi{} and CO emission of nearby face-on spiral galaxies established correlations between \rmol{} and other local quantities. \citet{Blitz2006} showed that \rmol{} has an almost linear dependence on midplane gas pressure, with significant variation between galaxies. \citet{Leroy2008,Eibensteiner2024} showed that \rmol{} correlates with gas surface density, midplane gas pressure, stellar mass surface density, and anti-correlates with galactocentric distance. They found typical conditions for the \hi{}-to-H$_2$ transition to be a gas weight $P_{\rm DE}\approx 2.3\times10^4 \ \rm K \ cm^{-3}$, gas surface density $\Sigma_{\rm gas}\approx 14 \ M_{\odot} \ \rm pc ^{-2}$, and stellar mass surface density $\Sigma_{\star}\approx 81 \ M_{\odot} \ \rm pc^{-2}$, all comparable but slightly higher than solar neighbourhood values ($\Sigma_{\rm gas}$ and $\Sigma_{\star}$ of $10$ and $40$ $M_{\odot} \ \rm pc^{-2}$, respectively).

Modeling the \hi{}-to-H$_2$ transition has been carried out using 1-dimensional, plane-parallel or spherical photodissociation region (PDR) models with varying levels of complexity \citep{Tielens1985,Black1987,LePetit2006,McKee2010,  Sternberg2014}. 

In recent years modeling of the star-forming ISM, including the \hi{}-to-H$_2$ transitions have also  been performed using 3-dimensional (magneto-)hydrodynamical simulations. The simulations range from molecular cloud scales \citep[e.g.,][]{Grudic2021,Grudic2022}, where cloud fragmentation and stellar accretion processes are resolved, to simulations of galaxies, either isolated \citep{Hu2017,Hu2019,Steinwandel2023}, or in mergers \citep{Lahen2019}, or in a cosmological context \citep{Hopkins2018}. 

While providing a more realistic large-scale context for ISM modeling, cosmological simulations must sacrifice spatial resolution and adopt sub-grid recipes to account for the unresolved substructure of the multiphase ISM (the exception being zoom-in simulations of dwarf galaxies, e.g., \citet{Gutcke2022a}). Galactic patch simulations \cite[e.g.,][]{Walch2015b,Iffrig2015,Kim2017,HSvD21,Gurman2025} aim to model an intermediate spatial scale, with a box size of $\sim$kpc, spatial resolution of $\lesssim \rm pc$, and run times of several 100 Myr. Implementation of star formation varies between models, but generally allows for a self-consistent treatment of feedback mechanisms such as supernova (SN) explosions, photoionizing radiation, and stellar winds, thanks to sufficient spatial resolution.

In one of the early versions of the SILCC project \citet{Walch2015b} compared $f_{\rm H_2}=M_{\rm H_2}/M_{\rm H}$ in a suite of simulations of solar neighbourhood conditions in a $0.5\times0.5\times5$ kpc$^3$ volume at 3.9 pc resolution. They included time-dependent thermochemistry, SN feedback, a constant background far-UV radiation field, a treatment for gas- and dust-radiation shielding, and ran their simulations for 100 Myr. By running simulations both with and without self-gravity, varying the SN rate, SN positioning scheme, background far-UV intensity, and inclusion of magnetic fields \citep[see also][]{Girichidis2018}, they reached instructive conclusions on the effects of different simulation setups on $f_{\rm H_2}$. They found that self-gravity increased $f_{\rm H_2}$ from $\sim0.1$ to $\sim0.4$, and that varying the SN rate from 5 to 45 Myr$^{-1}$ changed $f_{\rm H_2}$ from $\sim0.7$ to $<0.1$. \citet{Girichidis2018} found that including magnetic fields led to a decrease in $f_{\rm H_2}$ in the first 60 Myr of simulation time, and that a stronger initial magnetic field lead to lower $f_{\rm H_2}$, likely due to magnetic pressure delaying disk compression and thus the onset of H$_2$ formation. Importantly, positioning SNe at the global density peak rather than at random positions reduced $f_{\rm H_2}$ from $\sim0.4$ to $<0.05$. The former SN positioning scheme is perhaps more realistic, as SNe are expected to go off in molecular clouds following the death of the first massive stars to form therein. As such, this positioning scheme leads SNe to preferentially affect the pre-existing molecular gas in which they are positioned, thus dramatically reducing $f_{\rm H_2}$.

In \citet{Gatto2017}, the model for SN positioning in SILCC was improved by following the formation and accretion histories of sink particles, which represent star clusters. This allowed for a more self-consistent SN positioning when compared with \citet{Walch2015b}, but the overall $f_{\rm H_2}$ was found to be sensitive to the choice of density threshold, peaking at $\sim0.2$ and $0.5$ for a threshold of $10^2$ and $10^4\ \rm cm^{-3}$, respectively. In a more recent iteration of SILCC which includes a model for cosmic-ray propagation and models with varying metallicity, \citet{Brugaletta2025} report $f_{\rm H_2}=0.1$ for solar metallicity, when restricted to a distance of 250 pc from the midplane.

\citet{Gong2018,Gong2020} used the first version of the TIGRESS simulations \citep{Kim2017} to study the effects of metallicity and environment on CO-to-H$_2$ conversion factor $X_{\rm CO}$. They derived chemical abundances in their simulation in post-processing, which required the choice of a chemical evolution time as input for their integration of the equations of chemical evolution. For a chemical evolution time of 50 Myr, they found a time-averaged value of $\left<f_{\rm H_2}\right>=0.12$. For a chemical evolution time of 5 Myr, perhaps more representative of the dynamical time of the neutral gas in their simulations \citep[see Appendix A in ][]{HSvD21}, they found $f_{\rm H_2}=0.05$. The latest version of TIGRESS \citep[TIGRESS--NCR;][]{Kim2023} includes on-the-fly calculation of the molecular fraction, but has not reported on findings of $f_{\rm H_2}$. 

\textbf{An additional approach to modeling the star-forming ISM at high resolution is to simulate sequentially smaller volumes (i.e., zooming-in), increasing the resolution with each reduction of the simulation volume. The SILCC-ZOOM project \citep{Seifried2017, Seifried2018,Seifried2020} simulated sub-regions of the output from the SILCC simulations at up to 0.06 pc resolution. They simulated the sub-regions for 0.5-4.5 Myr, and found a peak value of $f_{\rm H_2}\approx0.9$. This result is expected as the sub-regions were chosen for their dense and highly molecular gas, and are not representative of the larger $\sim \rm kpc$ scale environment.} \citet{Smith2020} simulated a high-resolution ($\sim 0.1$ pc) galaxy patch by using a multi-stage refinement scheme. They simulated a galaxy-scale ISM structure at low resolution without self-gravity or star formation, and then added more physics while improving resolution and reducing the simulated volume, taking into account the large-scale gravitational potential and feedback from the lower-resolution volume that contains it. They reported a value of $f_{\rm H _2}=0.21$ for a single snapshot at their medium refinement level. However, this number included mass in young (age$\,<4\ \rm Myr$) sink particles that represent star clusters. At the refinement level for which they reported this value of \rmol{}, the sink formation density threshold was $574\ \rm cm^{-3}$, and the star formation efficiency at the sink level was assumed to be 0.02, effectively meaning that most of the gas above this density threshold was unaffected by pre-SN feedback. Excluding this gas, $f_{\rm H_2}$ dropped to 0.11.

\citet{HSvD21} simulated a $1\times1\times20\ \rm kpc^3$ box at $m_{\rm g}=1\ M_{\odot}$ resolution (effective spatial resolution of 0.2 pc), with time-dependent chemistry. In their model, single stars are formed from gas particles according to a Jeans mass threshold and are assigned a mass according to a stochastic sampling from an assumed \citet{Kroupa2002} initial mass function. In turn, this mass determined the photoionizing photon budget of the star and whether it sets off a SN explosion at the end of its life. They also included a global background far-UV field which scales with the instantaneous star formation rate (SFR), and a model for gas- and dust-shielding. They found \rmol{} $\approx 2\%$.

In addition to modeling the properties of a self-regulated ISM, Galactic patch simulations are also an ideal setup to make predictions for, or attempt to reproduce, emission line properties from cold atomic or molecular gas. They have been successful at reproducing the observed CO-to-H$_2$ conversion factor \citep{Gong2018,Gong2020,Seifried2020,Hu2022a}. At the same time, they have reproduced the relation between [C~\textsc{II}] 158 $\mu$m line emission $L_{\rm [C\,{II}]}$ and SFR \citep{Franeck2018,Ebagezio2022}. Even so, \citet{Gurman2024} showed that the predicted ratio of [C~\textsc{II}]-to-H$_2$ conversion factor was low compared with the observed value \citep[see, e.g.,][]{Zanella2018}. Combined, these two results provide further evidence for an underproduction of H$_2$ in galactic patch simulations.

In light of the apparent tension between observations and simulations, in this paper, we demonstrate the effects of several physical assumptions and model components in ISM modeling on \rmol{}. We use the GHOSDT simulations \citep{Gurman2025}, a suite of high-resolution magnetohydrodynamical simulations with star-by-star treatment of star formation, effective resolution of 0.05 pc (at the threshold for star formation; for our highest resolution runs). We run our models for 500 Myr, which allows us to average out temporal fluctuations in SFR and \rmol{}, and effectively erase the highly ordered and unrealistic initial conditions which dominate the simulation for the first $\sim 200$ Myr. In Section \ref{section: observations}, we present the observations we benchmark our results against. In Section \ref{section: methods}, we describe our numerical setup. In Section \ref{section: results}, we present our resulting \rmol{}. In Section \ref{section: clumping}, we discuss the effects of considering the sub-grid density structure. We summarize our results in Section \ref{sec: summary}.

\section{Observational References -- PHANGS-ALMA Sample}
\label{section: observations}
We consider observations from the Physics at High Angular resolution in Nearby Galaxies with Atacama Large Millimeter/submillimeter Array survey \citep[PHANGS--ALMA;][]{Leroy2021,Leroy2021b,Querejeta2024,Chiang2024}. We leverage the high-level data products from PHANGS--ALMA introduced in \citet{Sun2022,Sun2023}. These include 1.5~kpc resolution maps of 80 nearby star-forming galaxies of several quantities of interest. Maps of the H$_2$ surface density $\Sigma_{\rm H_2}$ are derived from CO(2-1) line emission observed with ALMA. Maps of the \hi{} surface density $\Sigma_{\ehiv{}}$ are derived from 21 cm emission gathered from various programs using the Very Large Array (VLA), Australia Telescsope Compact Array (ATCA), and others \cite[see][for full list]{Sun2022}. The stellar mass surface density $\Sigma_{\star}$ is derived from Wide-field Infrared Survey Explorer \citep[WISE;][]{Leroy2019} and Spitzer Space Telescope images \citep{Sheth2010}. Using these quantities and additional assumptions, the published data also includes the total gas weight estimator $P_{\rm DE}$ (see below).

\begin{figure}
	
	\centering
    \includegraphics[width=0.9\columnwidth]{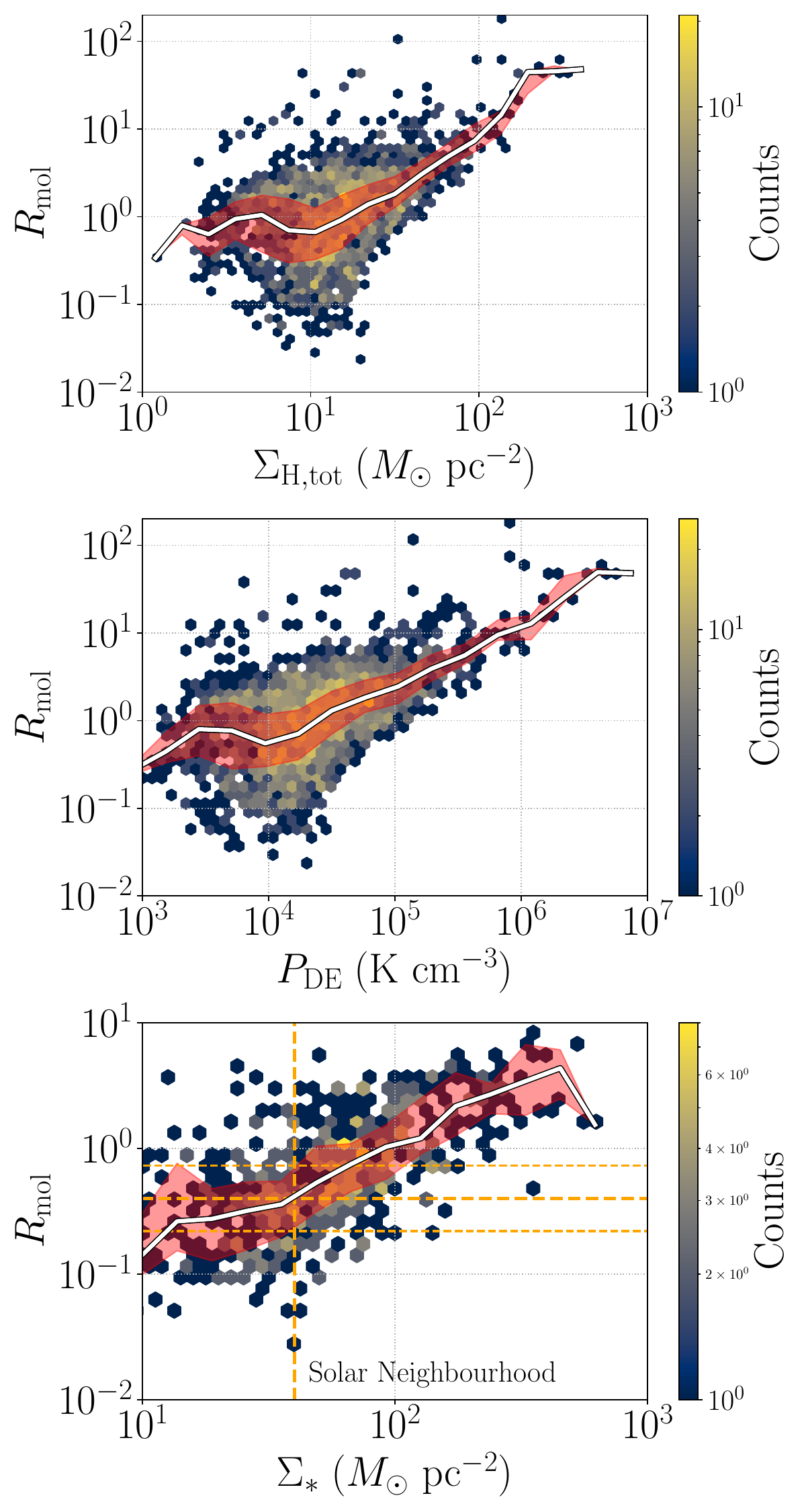} 
	\caption{
Observational data from the PHANGS--ALMA survey \citep{Leroy2021,Leroy2021b,Sun2022}. Top panel: $R_{\rm mol}$ as a function of total hydrogen surface density. Middle panel: $R_{\rm mol}$ as a function of gas weight. Bottom panel: $R_{\rm mol}$ as a function of stellar mass surface density for data with $\log \left(\Sigma_{\rm H ,tot}\right/\left( M_{\odot}\ \rm pc^{-2} \right))=1\pm 0.1$. In all panels, the white line shows the median and the red shaded region shows the 25-75\% range. The orange lines indicate the intersection of the median and interquartile range with the solar neighbourhood value of $\Sigma_{\star}=40\ M_{\odot}\ \rm pc^{-2}$.
}
		\label{fig: phangs data}
\end{figure}

We demonstrate the observed dependence of \rmol{} on the total (atomic+molecular) hydrogen surface density $\Sigma_{\rm H, tot}$ in the top panel of Figure \ref{fig: phangs data}. As the surface density increases, the gas density and the shielding from FUV radiation both increase, leading to an increase in \rmol{}. Due to a dominant contribution to the gravitational potential from the stellar component, the gas pressure is not determined by $\Sigma_{\rm H,tot}$ alone. Rather, it is determined by a combination of $\Sigma_{\rm H ,tot}$, the stellar (and dark matter) mass surface density, and the gas vertical scale height. We adopt the gas weight estimator $P_{\rm DE}$ \citep{Ostriker2010,Sun2020,Ostriker2022} which can be expressed as
\begin{equation}
    P_{\rm DE} = \frac{\pi G}{2} \,\Sigma_{\rm gas}^2+\Sigma_{\rm gas}\,\sqrt{2G\rho_{\star} }\,\sigma_{\mathrm{gas},z},
\end{equation}
where $\Sigma_{\rm gas}$ is the total gas surface density, $\rho_{\star}$ is the midplane stellar density, and $\sigma_{\mathrm{gas},z}$ is the velocity dispersion in the direction perpendicular to the disk plane. $\rho_{\star}$ is estimated using $\Sigma_{\star}$ and a radially varying disk thickness \citep{Kregel2002,Sun2020}. $\sigma_{\mathrm{gas},z}$ is assumed to be a constant $11\ \rm km\ s^{-1}$.

The middle panel of Figure \ref{fig: phangs data} shows \rmol{} as a function of $P_{\rm DE}$ in the PHANGS--ALMA data. We see a similar trend as with $\Sigma_{\rm H ,tot}$. This can be explained by the increased pressure due to increased gas weight leading to higher gas densities, increasing \rmol, but also by the expected correlation between the $\Sigma_{\rm H,tot}$ and $P_{\rm DE}$ \citep{Blitz2006}.

To perform a fair comparison for solar neighbourhood conditions, we examine a subset of the PHANGS--ALMA sample which has $\log \left(\Sigma_{\rm H,tot} / \left( M_{\odot}\ \rm pc^{-2}\right) \right)=1\pm0.1$. The bottom panel of Figure \ref{fig: phangs data} shows \rmol{} as a function of the stellar mass surface density for this subset of the data. We calculate a binned median and interquartile range for the subset, and find its intersection with the solar neighbourhood value of $\Sigma_{\star}=40\ M_{\odot}\, \rm pc^{-2}$. We find that it has a value of \rmol{} = 0.40, and its interquartile range spans $\left[ 0.22,\, 0.73\right]$. For the purpose of this work, we treat this value as the benchmark we wish to reproduce in our models, which assume the same stellar- and gas surface density\footnote{We note that there is a local minimum in the median observed \rmol{} dependence on $\Sigma_{\rm H,tot}$ (at $\sim10\ M_{\odot}\ \rm pc^{-2}$) and $P_{\rm DE}$ (at $\sim 10^4\ \rm K \ cm^{-3}$). This is likely explained by an observational bias.
In low surface density regions, the CO brightness at low \rmol{} can fall below the detection threshold, thus biasing the distribution toward high values of \rmol{}.
If this bias is real and present for the parameter space which we investigate, it could mean that the real value of \rmol{} is lower, perhaps closer to the range of 0.19-0.26 observed at the Solar circle.
}.

In the analysis that follows, we do not discuss the effects of the potential environmental or temporal variations in the CO-to-H$_2$ conversion factor $\alpha_{\rm CO}$ \citep[see, e.g.,][]{Bolatto2013, Gong2020,Hu2022a}. 

\section{Numerical Methods}
\label{section: methods}
Our simulation suite uses the numerical setup of the GHOSDT (Galaxy Hydrodynamical Simulations of a Supernova Driven Turbulent ISM) presented in \citet{Gurman2025} (hereafter \citetalias{Gurman2025}), an extension of the simulations of \citet{HSvD21} to include magnetic fields and varying gas surface density. For this work, we consider only models with $\Sigma_{\rm gas}=10\ M_{\odot}\ \rm pc^{-2}$ and solar metallicity, and vary several modeling choices and parameters as is described in Section \ref{section: sim list} below.

\subsection{Gravity and Magnetohydrodynamics}

The GHOSDT simulations are based on the publicly available version of GIZMO \citep{Hopkins2015}, a multimethod code implementing the meshless Godunov-type method \citep{Gaburov2011} on top of the TreeSPH code \textsc{GADGET}-3 \citep{Springel2005}. Gravity is solved using the ``treecode" method \citep{Barnes1986}. We treat MHD using the MFM method \citep{Hopkins2015}, and maintain $\nabla \cdot \bs{B}\approx0$ using a combination of 8-wave cleaning \citep{Powell1999} and the hyperbolic/parabolic divergence cleaning scheme presented in \citet{Dedner2002} following the updates from \citet{Tricco2012} and the (Riemann based) reformulation for the MFM method presented in \citet{Hopkins2016} (see \citetalias{Gurman2025} for more details).

\subsection{ISM Physics and Star Formation}

Following \citet{HSvD21}, we use a cooling and chemistry treatment based on \citet{Glover2007} and \citet{Glover2012}. We include time-dependent treatment of the chemistry of hydrogen species, combined with a HEALPIX \citep{Gorski2011} based method to calculate far-UV shielding. This includes shielding by both dust and H$_2$ from an effective shielding column density for each gas cell. \textbf{The chemical reaction network includes H$_2$ formation on dust, H$_2$ destruction by photodissociation, collisional dissociation and cosmic ray-ionization, and recombination in the gas phase and on dust grains. Cooling processes include fine-structure metal lines, molecular lines (including CO), Ly$\alpha$, H$_2$ collisional dissociation, collisional ionization of H, and recombination of H$^+$ in the gas phase and on grains. Heating processes include photoelectric emission from dust, cosmic-ray ionization, H$_2$ photodissociation, UV pumping of H$_2$, and H$_2$ formation. The unattenuated far-UV radiation field and the cosmic-ray ionization rate both scale with the star formation rate surface density.}

The model for star formation designates a gas particle as star forming when the local velocity divergence becomes negative and the thermal jeans mass $M_{\rm J}=\left(\pi^{2.5} c_s^3 \right)/\left( 6\,G^{1.5}\rho^{0.5} \right)$ drops below the kernel mass $M_{\rm ker}=N_{\rm ngb}\,m_{\rm g}$, where $c_s$ is the sound speed, $G$ is the gravitational constant, $\rho$ is the gas density, $m_{\rm g}$ is the gas particle mass, and $N_{\rm ngb}=32$ is the kernel size. This can be written as a condition on the temperature and density
\begin{equation}
\label{eq: sf prescription}
    n>1.4\times10^4 \ \mathrm{cm}^{-3} \left( \frac{T}{300} \right)^3\left(\frac{m_{\rm g}}{M_{\odot}}\right)^{-2}\left(\frac{\mu}{2.3}\right)^{-3}.
\end{equation}
This threshold is resolution-dependent by construction, as we expect to resolve higher density gas down to the Jeans scale as our resolution improves. \citet{HSvD21} showed that varying the resolution from 1 to 100 $M_{\odot}$ has essentially no effect on the time-averaged star formation rate, as it tends to self-regulate through photoionization and SN feedback. 

Once a gas particle is considered star-forming, it is assigned a probability of $\epsilon_{\rm sf} \Delta t/t_{\rm ff}$ to be converted into a star particle within a single time step, where $\Delta t$ is the time step, $t_{\rm ff}$ is the local freefall time, and $\epsilon_{\rm sf}$ is the star formation efficiency. We set $\epsilon_{\rm sf}=0.5$, which is a reasonable choice for a mass resolution of $1\ M_{\odot}$, for which the star formation threshold corresponds to a spatial resolution of 0.2 pc. While it might be an inadequate choice to keep $\epsilon_{\rm sf}$ constant for models with different resolutions, we choose to do so across all simulations considered in this work for simplicity. In addition to the criteria for star formation listed above, we enforce an instantaneous star formation density threshold of $n_{\rm isf}=10^5\ \rm cm^{-3}$. Above this density, a gas particle is converted to a star particle instantaneously.

Once formed, a star particle is stochastically assigned a mass by sampling a \citet{Kroupa2002} initial mass function (IMF). This mass is used to determine the photoionizing photon budget of the star and whether it ends its life in a SN explosion, but the star particle inherits the mass of its parent gas particle for the purpose of the gravitational field calculation. As stars form in large numbers in clusters, this method allows us to form stars one by one, predicting the right number of SNe per stars formed without violating mass conservation. If a star is assigned a mass $>8\ M_{\odot}$, it is designated as a massive star, and we treat its stellar feedback in the form of photoionization and SN explosions following \citet{Hu2017}. For each massive star, we determine the position of a quasi-spherical ionization front within which gas temperature is set to $10^4\ \rm K$ and hydrogen is assumed to be fully ionized. The radius of the ionization front is computed iteratively until the total recombination rate in the ionized region matches the mass-dependent ionizing photon budget of the star.

Massive stars also explode as SNe at the end of their mass-dependent lifetime. We model SNe by injecting $10^{51}\ \rm erg$ of thermal energy into the 100 nearest neighbours of the star. See \citetalias{Gurman2025}, \citet{Hu2019b}, and \citet{Steinwandel2020} for a discussion and justification of applying this model for SN feedback. We do not include a treatment for stellar winds, which can play a role in the evolution of molecular clouds and self-regulation of star formation \citep[see, e.g.,][]{Haid2018, Grudic2021,Grudic2022,Lancaster2021,Lancaster2021b}.

\subsection{Simulation Setup}

We simulate an elongated box with a spatial extent of 1 kpc in the $x$- and $y$- directions, and 100 kpc in the $z$ direction. Boundary conditions are periodic in $x$ and $y$, and outflow boundary conditions in $z$. We define the midplane as $z=0$, and $z=\pm50\ \rm kpc$ as the box boundaries. The initial gas temperature is set to $10^4\ \rm K$ and it is assumed to be entirely in the form of \hi{}. We adopt a constant solar chemical composition with a carbon and oxygen abundance of $1.4\times10^{-4}$ and $3.2\times 10^{-4}$, respectively \citep{Cardelli1996,Sembach2000}, and a dust-to-gas mass ratio of 0.01. Unless turned off, the initial magnetic field is set to $\bs{B}=\left( 5\, \mu \mathrm{G}\right)\, \hat{x}$ everywhere. The gas initially follows a vertical distribution of the form
\begin{equation}
    \rho \left( z \right) = \left(\frac{ \Sigma_{\rm g}}{2H_{\rm g}} \right) \mathrm{sech}^2 \left(\frac{z}{H_{\rm g}} \right),
\end{equation}
where $\Sigma_{\rm g}=10\ M_{\odot} \ \rm pc^{-2}$ is the gas surface density and $H_{\rm g}=250\ \rm pc$ is the initial disk scale height. This gives rise to an initial gravitational acceleration of the form
\begin{equation}
    a_{\rm g}=-2\pi G \Sigma_{\rm g} \mathrm{tanh} \left( \frac{z}{H_{\rm g}} \right).
\end{equation}
in the $z$-direction. Also included is a gravitational potential representing an old stellar population and a dark matter halo, whose dynamics we do not model explicitly. Acceleration due to the stellar component is assumed to take the form
\begin{equation}
    a_{\star}=-2\pi G \Sigma_{\star} \mathrm{tanh} \left( \frac{z}{H_{\star}} \right),
\end{equation}
where $\Sigma_{\star}=40\ M_{\odot}\ \rm pc^{-2}$ and $H_{\star}=250\ \rm pc$. The dark matter contribution is due to a Navarro-Frenk-White \citep{Navarro1997} profile with a virial mass $M_{\rm vir}=10^{12}\ M_{\odot}$ and concentration parameter $c=12$. The resulting acceleration is 
\begin{equation}
    a_{\rm DM} = -\frac{Gm\left( r\right)z}{r^3},
\end{equation}
where $r=\sqrt{z^2+R_0^2}$ represents the radial distance from the center of the halo, and $m\left(r\right)$ is the enclosed mass, given by
\begin{equation}
    m\left(r\right)=4\pi r_s^3\rho_s\ln\left[\left(1+r/r_s\right)-\left(r/r_s\right)\left(1+r/r_s\right)\right].
\end{equation}
In this expression, $r_s=17\ \rm kpc$, $\rho_s=9.5\times10^{-3}\ M_{\odot}\ \rm pc^{-3}$, and $R_0=8\ \rm kpc$ is the galactocentric distance to the Sun.

\subsection{Simulation List}
\label{section: sim list}
We run a set of variants on the simulations presented in \citetalias{Gurman2025}. We summarize the list of simulations and their associated names in Table \ref{table: simulation list}. In a first group of simulations, we aim to study the effect of numerical resolution. We run four simulations with the fiducial GHOSDT setup with varying particle mass, ranging from 100 to 0.25 $M_{\odot}$ (named m100 and m0.25, respectively). Assuming $T=100\ \rm K$, the corresponding spatial resolution (set by the SPH kernel length) at the density threshold for star formation \textbf{is 20, 2, 0.2, and 0.05 pc for models with $m_{\rm g}=100$, 10, 1, and 0.25 $M_{\odot}$, respectively}. For our best mass resolution, we scale down the box size in the $x$- and $y$- directions to 0.5 kpc in order to conserve computational resources. A second group of simulations is designed to investigate the effect of magnetic fields. We run pure-hydrodynamical simulations at both 10 and 1 $M_{\odot}$ resolution, labeled H-m10 and H-m1, respectively. The third group is aimed at testing for the effect of potential sub-grid density structure. We expand on this in Section \ref{subsec: sub-grid}. 

We group together a variety of simulations testing the effect of other parts of our simulation setup. We run these tests at $10\ M_{\odot}$ resolution \textbf{except for WSN-m1 which we run at $1\ M_{\odot}$ resolution}. In NPI-m10 we turn off photoionization feedback. In \textbf{WSN-m1 and} WSN-m10, we scale down the energy injected per supernova event by a factor of 3, to $3.33\times 10^{50}\ \rm erg$. L2-m10 has the box size in the $x$ and $y$ directions scaled up to 2 kpc. In LJ-m10 we introduce a variation on our star formation threshold, where we lower the right-hand side of Equation \ref{eq: sf prescription} by a factor of 100, to obtain the equivalent Jeans mass threshold for $1\ M_{\odot}$ resolution. While this allows gas to continue to gravitationally collapse to densities for which the Jeans mass is no longer resolved, we use this to complement our resolution study by degrading the mass resolution without changing the star formation prescription. Finally, in HG-m10 we scale up the stellar mass surface density in our external stellar gravitational potential from 40 to $400 \ M_{\odot}\ \rm pc^{-2}$. While technically not a solar neighbourhood model, we find it instructive to include this model in our study.

\begin{table*}

\centering
\begin{threeparttable}

\centering 
\begin{tabular}{l l l l}
\hline \hline 
\\ [-1.5ex] Group & Name & $m_{\rm g} \ \left( M_{\odot} \right)$ & Description \\ [0.5ex] 
\hline
\\[-3ex]Resolution Study & m0.25 & 0.25 & Box size scaled down by 2
\\[0.5ex]  & m1 & 1 &  Fiducial GHOSDT setup
\\[0.5ex]  & m10 & 10 &  
\\[0.5ex]  & m100 & 100 &  \\[0.5ex]
\hline
\\[-3ex] Magnetic Fields Off & H-m1 & 1 
\\[0.5ex]  & H-m10 & 10 &  \\[0.5ex]
\hline
\\[-3ex] Sub-grid Clumping & C4-m1 & 1 & H$_2$ formation artificially scaled up by 4
\\[0.5ex]  & FC-m1 & 1 & All 2-body chemical and collisional rates scale with $f_{\rm c}$
\\[0.5ex]  & FC-m10 & 10 & 
\\[0.5ex]  & FC-m100 & 100 & \\[0.5ex]
\hline
\\[-3ex] Variations & WSN-m1 & 1 & $E_{\rm SN}$ scaled down by 3
\\[0.5ex]  & WSN-m10 & 10 &  
\\[0.5ex]  & NPI-m10 & 10  & Photoionization switched off
\\[0.5ex]  & L2-m10 & 10 &  Box size scaled up by 2
\\[0.5ex]  & LJ-m10 & 10 &  Modified star formation prescription
\\[0.5ex]  & HG-m10 & 10 & $\Sigma_{\star}$ scaled up by 10\\[0.5ex]
\hline

\end{tabular}

\end{threeparttable}
\caption{Overview of our suite of simulations.}
\label{table: simulation list}
\end{table*}

\subsection{Model for Unresolved Density Substructure}
\label{subsec: sub-grid}
In models C4-m1 and FC-m1/10/100, we aim to investigate the effect of accounting for unresolved density substructure on \rmol{}. Our approach includes the introduction of a clumping factor, denoted as $f_c$, to account for the expected increase in the rate of two-body reaction rates in the case of unresolved clumpy structure. This builds on a similar approach implemented by \citet{Lupi2018} in simulations of an isolated galaxy, albeit at lower resolution. The clumping factor is calculated by assuming an unresolved sub-grid log-normal density distribution due to supersonic turbulent motions. 

Numerical experiments of artificially driven supersonic turbulence using idealized hydrodynamical simulations \citep{Padoan1997,Passot1998} find that the resulting log-normal probability distribution function has a (logarithmic) standard deviation of $1+\left(b\, \mathcal{M}\right)^2$, where $\mathcal{M}$ is the Mach number, and $b$ is a prefactor calibrated numerically. We define the clumping factor 
\begin{equation}
    f_c\equiv1+\left(b\, \mathcal{M}\right)^2.
    \label{eq: clumping def}
\end{equation}
In addition, for a log-normal PDF, it follows that
\begin{equation}
    f_c=\frac{\left< n ^2 \right>}{\left< n \right>^2}.
\end{equation}
Equivalently, $f_c$ quantifies the enhancement in two-body reaction rates compared to the constant density case. \citet{Federrath2008} found that $b\approx 1$ for purely compressive forcing, $b\approx 1/3$ for purely solenoidal forcing, and $b\approx 0.5$ for a 50-50 mix of the two. In what follows, we adopt $b=0.5$. This sub-grid description should not be considered definitive, due to this assumption of constant $b$, omission of intermittency, and omission of physics other than isothermal turbulence. Nevertheless we expect it to model the first-order effect upon the clumping factor: the presence of unresolved structures formed by $\propto \mathcal{M}^2$ isothermal compressions.

We estimate $f_c$ following treatment in the FIRE-3 model \citep{Hopkins2023}. The Mach number is calculated using the ratio of the thermal velocity $v_{\rm therm}=\sqrt{3k_B T/\mu \, m_p}$ and the turbulent velocity $v_{\rm turb}\equiv h\left|\nabla v\right|$, where $\mu$ is the mean molecular weight, $h$ is the cell size and $\nabla v$ is the local velocity gradient. The H$_2$ abundance weighted average of $f_c$ taken over snapshots in the time range 200-500 Myr of model m1 is $\approx4$. This is already an indication that sub-grid clumping is expected to play a non-negligible role in the gas chemistry, an aspect usually ignored in simulations at this resolution. Motivated by this, in model C4-m1, we apply an ad-hoc assumption of $f_c=4$, and apply it to H$_2$ formation on dust grains only. Thus, we adjust the formation rate coefficient of H$_2$ on dust grains
\begin{equation}
    R_{\rm H _2}=\frac{3.0\times 10^{-17}\sqrt{T_2} }{1+0.4\sqrt{T_2+0.15}+0.2T_2+0.08T_2^2}\ \rm cm^{-3}\ s ^{-1},
\end{equation}
where $T_2=T/100\ \rm K$, up by a factor of 4. 

In models FC-m1/10/100, we calculate $f_c$ on the fly and scale up all two-body collisional and chemical reactions in our thermochemistry network in each cell by a factor of $f_c$. This scaling up of two-body reaction rates is only valid under the assumption that the mixing time for gas within a given cell is shorter than the characteristic timescale for the reactions of interest. 
We demonstrate the validity of this assumption in Section \ref{section: clumping}.

\section{Results}
\label{section: results}
\subsection{Overview}
In Figure \ref{fig: h2 results}, we present the time dependence of \rmol{} across our different models. All models show significant temporal fluctuations in \rmol. The peaks in \rmol{} correlate with the peaks in SFR due to a large fraction of the gas residing in dense and shielded regions \citep[see, e.g.,][]{Gurman2024}. These fluctuations follow the cycle of formation of cold dense gas clouds, gravitational collapse, a rise in SFR, and subsequent heating, ionization, and dispersal of the cloud by stellar feedback. We summarize the time-averaged star formation rates and \rmol{} in 
panel (f) of Figure \ref{fig: h2 results}.
With the exception of model HG-m10, the SFR surface density falls within the range 
$10^{-3.02}$ to $10^{-2.62}$$\ M_{\odot} \ \rm yr^{-1} \ kpc^{-2}$.
Given the strong temporal fluctuations in SFR in all of these models, this variation of 0.4 dex can be considered a good agreement between the different models. The elevated SFR in HG-m10 of $10^{-2.19}\ M_{\odot} \ \rm yr^{-1} \ kpc^{-2}$  is expected as the enhanced background stellar potential increases the gas weight, which is subsequently balanced by increased pressure driven by increased SFR \citep{Ostriker2022}.

\subsubsection{Resolution Dependence}
The effect of resolution on \rmol{} is demonstrated in panel (a) of Figure \ref{fig: h2 results}. We find an increase in \rmol{} of a factor of 3.3 and 2.5 when varying the resolution from 100 to 10, and from 10 to 1 $M_{\odot}$, respectively. The change from 1 to 0.25 $M_{\odot}$ is only a factor of 1.2, indicating that we may be approaching convergence. The resolution dependence of \rmol{} is caused by two factors. First, as we increase resolution, we resolve higher-density gas. This is made clear by our Jeans mass threshold for star formation shown in Equation \ref{eq: sf prescription}. This means that the cold neutral medium (CNM) is not resolved at $m_{\rm g}=100\ M_{\odot}$, and barely resolved at $m_{\rm g}=10\ M_{\odot}$. This affects the H$_2$ mass that resides in the CNM. Even so, \rmol{} in model LJ-m10 (shown in panel (c) of Figure \ref{fig: h2 results}), which uses the same Jeans mass threshold as m1, agrees with that of model m10. Therefore, it is not our star formation recipe that drives the increase in \rmol{} between models m10 and m1. Instead, this increase is rather due to an improved resolution of the clumpy substructure of the ISM. 

\textbf{This resolution effect has been demonstrated by \citet{Seifried2017}, who experimented with varying the maximum refinement level in their adaptive mesh refinement (AMR) scheme. They simulated two sub-regions of the SILCC simulations \citep{Walch2015,Girichidis2016} which contained a molecular cloud, and found that \rmol{} at a given simulation time increased with improved resolution, and converged when the maximum refinement level reached a resolution of 0.24 pc or better. Using the same treatment for hydrodynamics, chemistry, and radiation shielding, \citet{Joshi2019} tested the resolution dependence of \rmol{} in simulations of a turbulent driven box and colliding flows, both without self-consistent star formation or feedback. They found that a maximal refinement level better than 0.25 pc was needed for convergence of \rmol{}. Our result that \rmol{} approaches convergence at $1\ M_{\odot}$ resolution (equivalent to $\sim0.2$ pc spatial resolution) is consistent with their findings. We discuss the resolution requirement proposed by \citet{Joshi2019} further in the context of this work in Section \ref{section: clumping}}. 

\subsubsection{Effect of Magnetic Fields}
Panel (b) of Figure \ref{fig: h2 results} demonstrates the effects of magnetic fields on \rmol{} by comparing MHD and pure-hydrodynamical runs for mass resolution 10 and 1 $M_{\odot}$. Including magnetic fields leads to an increase in \rmol{} of a factor of 1.5 and 3.5 for 10 and 1 $M_{\odot}$, respectively. As discussed in \citetalias{Gurman2025}, magnetic pressure reduces fluctuations in the SFR, while keeping the time-averaged SFR largely unchanged. As a result, the effective disk-height decreases and the cold gas mass fraction increases, which explains the increase in \rmol{}. Even so, \rmol{} in the fiducial solar neighbourhood model m1 still falls short of the observed value by a factor of $\sim4$.
Our results are consistent with \cite{HSvD21}, who used pure-hydrodynamical simulations and an otherwise identical simulation setup and found that \rmol{} converges at $m_{\rm g}\leq10 \ M_{\odot}$.
Since they did not include magnetic fields in any of their models, this suggests that capturing the effect of magnetic fields requires better resolution. It is possible that due to magnetic energy being distributed from small to large scales, the effects of magnetic pressure are sensitive to numerical resolution.

\subsubsection{Variations in Simulation Setup and Feedback}

Model HG-m10, presented in panel (c) of Figure \ref{fig: h2 results}, shows an increase of a factor of $2.0$  in \rmol{} when compared with m10. An increase in \rmol{} is expected when $\Sigma_{\star}$ increases as the increased background gravitational field drives gas pressure up. Even so, the observational data presented in the bottom panel of Figure \ref{fig: phangs data} suggest a steeper dependence of \rmol{} on $\Sigma_{\star}$.

Our experiments with variations on feedback models are shown in panel (d). In \textbf{WSN-m1 and} WSN-m10, we reduce the energy injected into the ISM per SN event by a factor of 3. \textbf{This results in an insignificant increase by a factor of 1.15 (WSN-m10) and 0.9 (WSN-m1) in \rmol{} when compared with models m10 and m1, respectively}. The lower SN energy is still sufficient to disperse molecular clouds, and therefore the effect on the gas conditions is minor. Even so, there is a marginally significant increase in SFR, likely due to a lower turbulent feedback yield when SNe are weakened. In model NPI-m10, where we turn off photoionization feedback, we find a factor of $2.0$ reduction in \rmol{} compared with m10. While perhaps counterintuitive at first, this is explained by the fact that photoionization feedback, as our only source of pre-SN feedback, plays an important role in lowering the clustering of SNe, which, in turn, lowers the effectiveness of SNe in disrupting the disk \citep{Smith2021,Hu2023b}. \textbf{Models LJ-m10 and L2-m10 do not show a significant effect on \rmol{} when compared with m10.}

\subsubsection{Effect of Sub-Grid Clumping}

\begin{figure*}
	
	\centering
	\includegraphics[width=1.9\columnwidth]{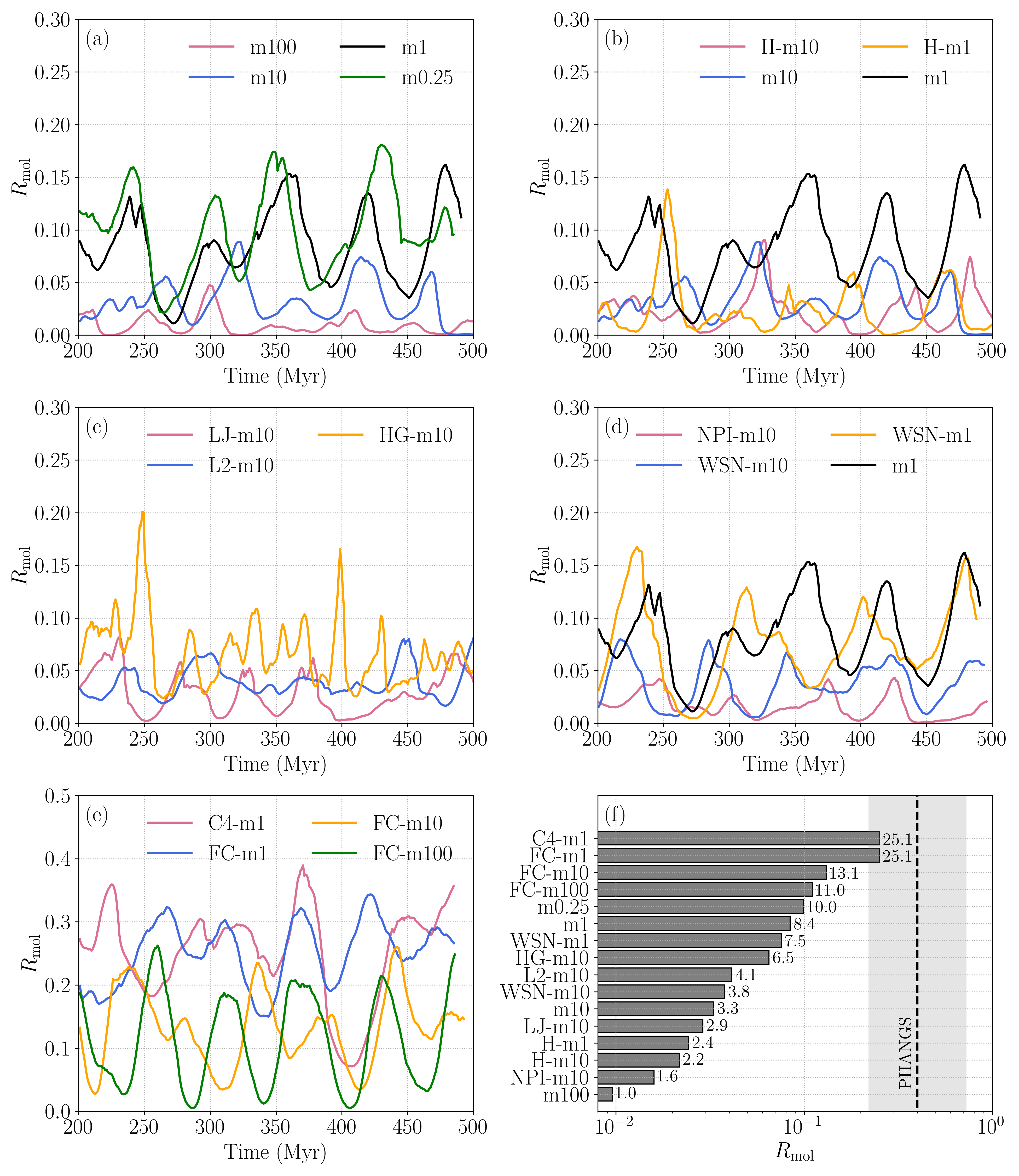} 
	\caption{
(a)-(e): The H$_2$-to-\hi{} mass ratio \rmol{} as a function of time for different simulation groups. (a): Resolution study. (b): Effect of magnetic fields. (c)-(d): Variations to feedback and simulations setup. (e): Effects of our clumping sub-grid model. (f): Time averaged values of \rmol{} for all simulations. Vertical dashed line and shaded region show the observed median and interquartile range for PHANGS-ALMA pixels with $\Sigma_{\rm H,tot}=10$ and $\Sigma_{\star}=40$ $M_{\odot}\ \rm pc^{-2}$. Numbers indicate \rmol{} in \% for each model.
}
		\label{fig: h2 results}
\end{figure*}

We present models C4-m1 and FC-m1/10/100 in panel (e) of Figure \ref{fig: h2 results}. With these models, we attempt to account for unresolved substructure due to supersonic gas motions. In model C4-m1, we scale up the H$_2$ rate coefficient for formation on dust grains by a factor of 4. This number is the result of computing the clumping factor $f_c$ using Equation \ref{eq: clumping def} for each particle in each snapshot in m1 and taking the H$_2$ mass-weighted average. Model C4-m1 incorporates a rather ad-hoc modification to our chemical model, and does so inconsistently, as we apply a clumping factor to H$_2$ formation only rather than to all two-body reactions, both chemical and radiative. In addition, the enhancement in H$_2$ formation is applied independent of gas density or temperature. Although expected, the result of an increase of a factor of $3.0$ is promising as it drives \rmol{} closer to the observed value. 

In models FC-m1/10/100, we implement a self-consistent sub-grid model accounting for the unresolved density substructure. Here, $f_c$ is calculated on the fly for each particle and enters our network of cooling and chemical reactions by multiplying the reaction rate of every two-body reaction in our network. The time averaged, H$_2$ mass-weighted average $f_c$ for models m100, m10, and m1 is 13, 6, and 4, and we find an increase in \rmol{} by a factor of 11.0, 3.97, and 3.0 in models FC-m100, FC-m10, and FC-m1, respectively. \textbf{The striking increase in \rmol{} between models m100 and FC-m100 is due to the resolution dependence of $f_c$, as we discuss further in Section \ref{section: clumping}.} Our most promising candidate for explaining the discrepancy between models and observations is model FC-m1, in which \rmol{}  falls within the inter-quartile range of the observed \rmol{} at solar neighbourhood conditions, and in agreement with measurements in the Solar circle. This demonstrates that unresolved substructure is important for H$_2$ formation in the ISM even at a mass resolution of 1 $M_{\odot}$. 

\subsection{The Importance of Sub-Grid Structure}
\label{section: clumping}
In this section, we explain how and why accounting for unresolved density substructure affects \rmol{}. The result that \rmol{} increases when scaling up collisional rates by a factor $> 1$ is expected. The formation of H$_2$ is dominated by a collisional process scales as $n^2$ and is locally enhanced by a factor of $f_c$. H$_2$ destruction is dominated by photodissociation whose rate does not explicitly depend on the the assumption of unresolved substructure, as the mean density within a gas cell remains unchanged. An indirect effect does enter through the effect on the H$_2$ shielding column, which will be higher, on average, at a given density when H$_2$ formation is enhanced. Another effect on shielding, which we do not capture in our sub-grid treatment, is higher porosity of the gas to far-UV radiation.

\subsubsection{Comparison with Fiducial Model}
\begin{figure}
	
	\centering
	\includegraphics[width=1\columnwidth]{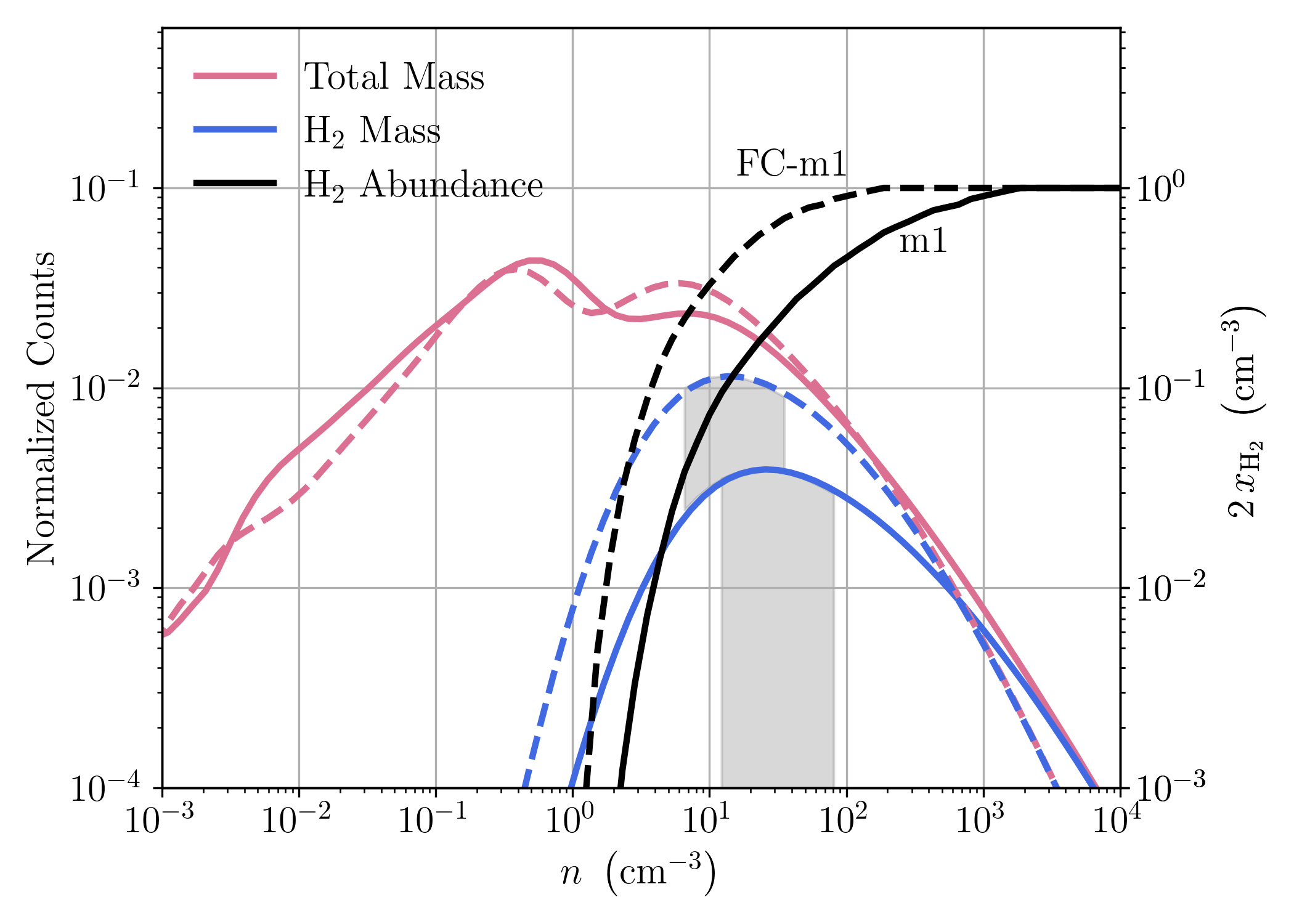} 
	\caption{
Normalized histogram of the total gas mass (violet) and H$_2$ mass (blue) in model m1 (solid lines) and FC-m1 (dashed lines). The black solid and dashed lines show the binned median H$_2$ mass fraction as a function of density for the same models. The shaded regions indicate the density range containing the 25-75\% range of H$_2$ mass in the respective models.
}
		\label{fig: h2 mass hist}
\end{figure}

Most of the H$_2$ mass to be gained by accounting for sub-grid clumping is from the lower density CNM. Approximately half of the H$_2$ mass is in the partially molecular, but much more abundant CNM gas with $n\sim 10-100\ \rm cm ^{-3}$, as opposed to the fully molecular gas with $n\gtrsim10^3\ \rm cm^{-3}$. Figure \ref{fig: h2 mass hist} shows the normalized mass- and H$_2$ mass-weighted histograms, and the median H$_2$ abundance as a function of density for models m1 (solid lines) and FC-m1 (dashed lines). While \hi{} in model m1 is fully converted to H$_2$ only at $n\sim 10^3\ \rm cm^{-3}$, the wide and prominent peak at lower densities in the mass-weighted histogram leads to 50\% of H$_2$ mass residing in the density range 10-100 $\rm cm ^{-3}$. When the sub-grid model is applied, the total density PDF does not change drastically, but the molecular abundance at a fixed density does increase significantly. \textbf{Sub-grid clumping has negligible effect on the H$_2$ mass in the high density tail of the PDF, as gas at this density is fully converted to H$_2$ even in the absence of sub-grid clumping.} While the H$_2$ mass in the high-density tail of the PDF does not change much, the increased H$_2$ abundance in the partially molecular gas drives the large increase in total H$_2$ mass.

We note that this intermediate density range, which is only partially molecular but dominates the H$_2$ mass budget, is expected to display significant sub-grid clumping. 
Due to the Lagrangian nature of {\sc Gizmo},
our spatial resolution of 0.2 pc $\times \left( m_{\rm g} / M_{\odot} \right)^{1/3}$ at $n_{\rm H}=1.4\times 10^4 \ \rm cm^{-3}$ (the typical density for star formation in our m1 model) scales to 2 pc at $n_{\rm H}\sim 10 \ \rm cm^{-3}$, leading to more unresolved substructure and thus a non-negligible $f_c$.

\subsubsection{Resolution and Density Dependence of $f_c$}
\begin{figure}
	
	\centering
	\includegraphics[width=0.9\columnwidth]{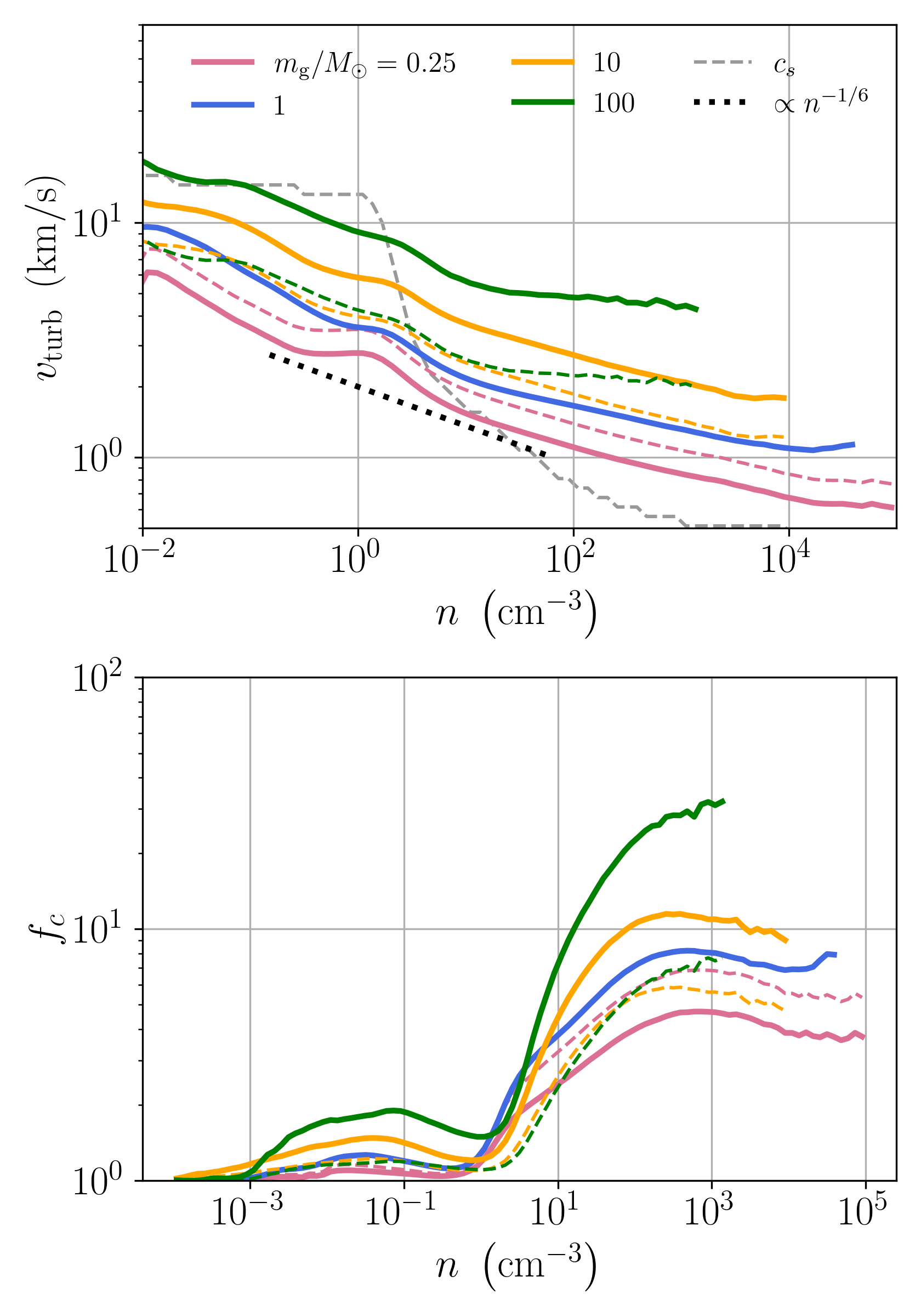} 
	\caption{
Top panel: $v_{\rm turb}$ as a function of density for models m100, m10, m1, and m0.25. Dashed colored lines show the corresponding solid curves scaled by $\left(m_{\rm g}/M_{\odot}\right)^{-1/6}$. Also shown is the sound speed for m1, which is essentially independent of resolution, and a power law of the form $n^{-1/6}$ for reference. Bottom panel: as top but for $f_c$. The dashed lines show the solid curves scaled according to Equation \ref{eq: clump scale}.
}
		\label{fig: v_turb f_c}
\end{figure}

The top panel of Figure \ref{fig: v_turb f_c} shows the mean sub-grid turbulent velocity as a function of density. \textbf{If we assume a size-linewidth relation of the form}
\begin{equation}
 v_{\rm turb} \propto h^{1/2}
\end{equation}

\textbf{\citep{Larson1981}, where $h$ is the cell size, we can expect a relation between the turbulent velocity $v_{\rm turb}$ and the particle mass $m_{\rm g}$ of the form}
\begin{equation}
    v_{\rm turb}\propto n^{-1/6}.
\end{equation}
\textbf{This is due to the fact that in a given simulation our particle mass is constant and therefore 
\begin{equation}
    h\propto \left(m_{\rm g} / n\right) ^{1/3}.     
\end{equation}
Equivalently, at a fixed density, we would expect the scaling of $v_{\rm turb}$ with particle mass to follow}
\begin{equation}
    v_{\rm turb}\propto m_{\rm g}^{1/6}.
\end{equation}

\textbf{To demonstrate the consistency of our model with this assumption,} we also plot the same curves where $v_{\rm turb}$ is divided by $\left(m_{\rm g}/M_{\odot}\right)^{1/6}$. The curves follow a similar density dependence of roughly $\propto n^{-1/6}$, and once the resolution dependence is factored out agree in value as well. The computed turbulent velocity is also in rough agreement with the observed \citet{Larson1981} size-linewidth relation, which is $\approx 1\ \rm km\ s^{-1}$ at a scale of 1 pc. For $m_{\rm g}=1\ M_{\odot}$, 1 pc corresponds to $n\approx13\ \rm cm ^{-3}$, for which we find $v_{\rm turb}=2.5\ \rm km\ s ^{-1}$. At densities above the star formation threshold, which is resolution dependent, the curves diverge. 

The bottom panel of Figure \ref{fig: v_turb f_c} shows the average $f_c$ as a function of density.  As the sound speed, shown in the top panel of Figure \ref{fig: v_turb f_c}, drops steeply at $n \sim 5 \ \rm cm^{-3}$ due to the collisional excitation of fine structure transitions, $f_c$ increases. At higher densities, the density dependence of the temperature weakens, and we expect the decrease in $v_{\rm turb}$ to drive $f_c$ to lower values, but this density regime is only marginally resolved in the m0.25 model, where we observe a shallow decline of $f_c$ at the highest densities. Following the same line of thought applied to the resolution dependence of $v_{\rm turb}$ \textbf{and using Equation \ref{eq: clumping def}}, we can deduce that at a given temperature $\mathcal{M}\propto v_{\rm turb}$ and therefore
\begin{equation}
    f_c-1=\left(b\,\mathcal{M}\right)^2\propto v_{\rm turb}^2 \propto m_{\rm g}^{1/3}.
\end{equation}
This resolution dependence is demonstrated with the dashed lines in the bottom panel of Figure \ref{fig: v_turb f_c}, where instead of $f_c$ we plot
\begin{equation}
\label{eq: clump scale}
    \widetilde{f}_{c}\equiv\left(f_c-1\right)\times \left(m_{\rm g}/M_{\odot} \right)^{-1/3}+1,
\end{equation}
and find a reasonable agreement between the different dashed curves and the m1 curve, indicating that the scaling with resolution is well understood. This analysis shows that at a spatial resolution of $\approx2$ pc (the spatial resolution of our $m_{\rm g}=1\ M_{\odot}$ models at $n=10\ \rm cm^{-3}$) there is sufficient unresolved substructure which requires sub-grid modeling in order to capture H$_2$ formation correctly.

\subsubsection{Sub-Grid Model Applicability}
\begin{figure}
	
	\centering
	\includegraphics[width=1\columnwidth]{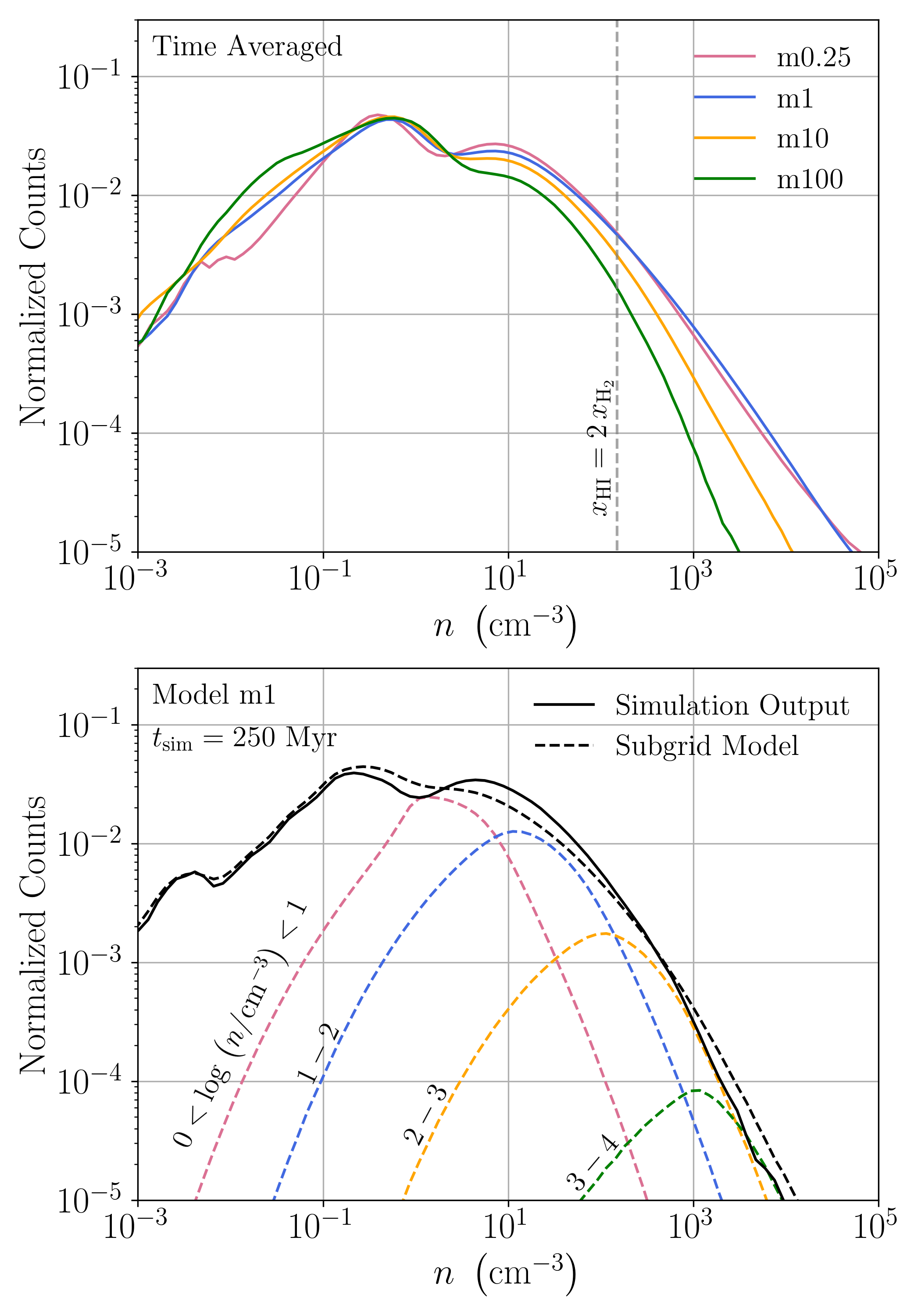} 
	\caption{
Top panel: normalized mass-weighted density histograms of the total gas mass for simulations in our resolution study, averaged over simulation time 200-500 Myr. The dashed vertical line marks the \hi{}-to-H$_2$ transition density for our m1 model. Bottom panel: mass-weighted density histogram (solid black) and the modified histogram resulting from our sub-grid model (dashed black; see Section \ref{section: clumping}), both for model m1. Colored dashed lines show the modified histogram when applied to density bins with a width of 1 dex.
}
		\label{fig: convergence test}
\end{figure}

Finally, we address the application of a sub-grid model in a seemingly converged model. In the top panel of Figure \ref{fig: convergence test}, we show the time-averaged density histograms of models m100, m10, m1, and m0.25. There seems to be little change in the high-density tail between models m1 and m0.25, consistent with there being a small difference in \rmol{} between the two. Thus, the question arises whether this convergence of the density PDF contradicts the result of $f_c>1$ and the subsequent increase in \rmol{} when it is applied. To further challenge this, we demonstrate the density histogram of a single snapshot from model m1 alongside the expected density histogram if the sub-grid clumping were resolved. We compute this by drawing 10 new particles for each particle in the snapshot output. We draw these particles from a density PDF corresponding to the computed $f_c$ of each particle and reconstruct the density histogram, after having essentially sampled the sub-grid structure of unresolved gas cells. The new density histogram is not significantly different from that of the original snapshot. The overall shape of the distribution is maintained, no large amount of gas is moved from low-density bins to higher ones, and only a small mass of gas is added at the high-density tail, which constitutes only a small fraction of the total H$_2$ mass. We also break the resampled density histogram into several components, which originate in particles within density bins with a width of 1 dex. It is clear that while the $\log \left( n/\rm cm ^{-3}\right)=1-2$ bin obtains a high-density tail, the mass in that tail is completely negligible compared with the total mass in the snapshot in that density range. 

Considering the aforementioned analysis, we note that the application of our sub-grid clumping model accounts for an enhancement in $\left<n^2\right>$ irrespective of actually adding any dense gas. Rather, the dense unresolved clumps in which formation is enhanced within a larger gas cell are assumed to mix quickly in with the rest of the density distribution, thus enhancing \rmol{}. Since this enhancement is due to the increase in $x_{\rm H_2}$ in the 10-100 cm$^{3}$ gas, it is significant even when the density PDF is converged and the characteristic density for the \hi{}-to-H$_2$ transition of $\approx 10^{3}\ \rm cm^{-3}$ is resolved.

For the assumption of fast mixing to hold, the timescale for gas mixing within a given gas cell must be short compared to other relevant timescales, in this case the H$_2$ formation timescale $t_{\rm H_2}$.
We approximate an upper bound on the gas mixing time using the local crossing time, which we define as $t_{\rm cross,local}\equiv h/v_{\rm turb}$ where $h$ is the cell size and $v_{\rm turb}$ is the turbulent velocity we describe in Section \ref{section: methods}. \textbf{The condition for the validity of our sub-grid model is therefore}
\begin{equation}
    t_{\rm cross,local}<t_{\rm H_2}.
\end{equation}
Using the results shown in Figure \ref{fig: v_turb f_c}, we estimate 
\begin{equation}
    v_{\rm turb}=4\ \mathrm{km\ s^{-1}}\left(m_{\rm g}/M_{\odot}\right)^{1/6}\left(n/\mathrm{cm}^{-3}\right)^{-1/6}
\end{equation}
and the corresponding local crossing time
\begin{equation}
    t_{\mathrm{cross,local}}=\frac{h}{v_{\mathrm{turb}}}=1.05\ \mathrm{Myr}\,\left(m_{\mathrm{g}}/M_{\odot}\right)^{1/6}\left(n/\mathrm{cm}^{-3}\right)^{-1/6}
\end{equation}
At the same time, for a temperature of $100\ \rm K$ we can estimate the H$_2$ formation time
\begin{equation}
    t_{\mathrm{H}_2}=\frac{1}{2\,R_d\,n}\approx 1.5 \,\mathrm{Gyr} \,\left( n/\mathrm{cm}^{-3}\right)^{-1}.
\end{equation}
We thus obtain the ratio of the mixing time and the H$_2$ formation time
\begin{equation}
    \frac{t_{\mathrm{H}_2}}{t_{\mathrm{cross,local}}}=1.4\times 10^{3}\left(n/\mathrm{cm}^{-3}\right)^{-5/6}\left(m_{\mathrm{g}}/M_{\odot}\right)^{-1/6}.
\end{equation}
For the worst case of $m_{\rm g}=100\ M_{\odot}$, we find that $t_{\rm H_2}=t_{\rm cross,local}$ for $n\approx 3000 \ \rm{cm}^{-3}$. This is above the density at which hydrogen in our simulations is fully converted into molecular form. Therefore, the assumption of rapid mixing is valid for the density range of interest in which $2x_{\rm H _2}<1$. We note, however, that other thermochemical processes might be affected by applying our sub-grid model at densities for which the characteristic timescale $\tau$ is shorter than the cell crossing time, and a more careful application of the model should be considered, e.g., assuming that $f_c=1$ for cells where $\tau>t_{\rm cross,local}$.

\textbf{The condition for the validity of our sub-grid model on the H$_2$ formation timescale is different from the criterion suggested by \citet{Joshi2019} for the convergence of \rmol{} in their simulations. They demonstrated that \rmol{} converges once the resolution meets two conditions. First, it must be sufficient to resolve the density at which the gas becomes fully molecular. The corresponding density which must be resolved is in the range 1-10 cm$^{-3}$, depending on their simulation setup. Our simulations meet this condition for $m_{\rm g}\leq 10\ M_{\odot}$ \citep{HSvD21}. Second, they require that 
\begin{equation}
    t_{\rm H_2}<t_{\rm cross, glob},
\end{equation}
where $t_{\rm cross,glob}$ is the global cell crossing time due to bulk motion. It is defined as $t_{\rm cross,glob}=h/v_{\rm rms}$, where $v_{\rm rms}$ is the mass-weighted root-mean-square velocity in the simulation. This resolution criterion ensures that gas can spend enough time in a high-density region and form H$_2$ before it moves to a low-density region. Similarly, the condition for our sub-grid model applicability $t_{\rm H_2}>t_{\rm cross,local}$, where $t_{\rm cross,local}=h/v_{\rm turb}$ (i.e., the sub-grid crossing time) ensures that sub-grid mixing is fast enough to enhance the H$_2$ abundance in a given cell.}

\textbf{Because the H$_2$ mass in the simulations of \citet{Joshi2019} was dominated by gas cells with a molecular fraction $2x_{\rm H_2}\approx1$, their resolution criteria were sufficient. As we have discussed, accounting for sub-grid clumping in fully molecular gas has no effect on the resulting molecular fraction. In our simulations, however, a significant fraction of the H$_2$ mass resides in predominantly atomic gas. This is due to dynamical effects and the effect of stellar feedback, which are not accounted for in \citet{Joshi2019}. In this case, the resolution criterion of \citet{Joshi2019} is insufficient, because it does not account for the need to capture the molecular abundance in partially molecular and less dense gas. As we have demonstrated, a stricter resolution requirement arises in this case, and the sonic scale of the cold ISM \citep[$\sim 0.2$ pc; ][]{Xu2019} would have to be resolved at all densities where molecular gas is present in order for \rmol{} to be computed correctly.}

\section{Summary}
\label{sec: summary}

We have presented a set of simulations of a self-regulated, star-forming ISM demonstrating the effect of different modeling choices on the molecular hydrogen fraction \rmol{}. This has been done in light of an apparent discrepancy of a factor of $\sim4$ between the observed value of \rmol{} and resolved ISM simulation of $\sim \rm kpc$ scales. We explored, among other things, the effect of resolution, inclusion of magnetic fields, and a model for unresolved density substructure. 

\begin{enumerate}[leftmargin=*, noitemsep]
    \item Our density PDF and \rmol{} show signs of likely convergence at $m_{\rm g}=1\ M_{\odot}$, with a factor of $\sim 3$ increase in \rmol{} going from 100 to 10 to 1 $M_{\odot}$.
    \item Including magnetic fields leads to a factor 3.5 increase in \rmol{} for $m_{\rm g}=1\ M_{\odot}$, likely due to a modulation of SFR \citepalias{Gurman2025}.
    \item Excluding photoionzation lowers \rmol{} by a factor of 2 for $m_{\rm g}=10\ M_{\odot}$, likely due to more clustered star formation and SN feedback \citep[see][]{Hu2022a}.
    \item \rmol{} is insensitive to lowering of SN energy by a factor of 3, increasing simulation volume, and varying the star formation Jeans mass threshold.
    \item Accounting for sub-grid clumping gives a factor of $\sim 3$ increase in \rmol{} at $m_{\rm g}=1 \ M_{\odot}$, due to an enhanced H$_2$ formation rate in intermediate density gas with $n\sim 10-100\ \rm{cm}^{-3}$. We point to this model as our most successful effort to relieve the tension between the simulated and observed values of \rmol{}.
    \item We compared our predictions for \rmol{} by directly measuring the H$_2$ content in our simulations, while observations measure $\Sigma_{\rm H_2}$ indirectly using CO line emission. We point to post processing our simulations to predict CO line luminosity and make a direct comparison with observations as a prospect for further study.

\end{enumerate}

This work highlights the importance of using sufficiently high resolution, including pre-SN feedback and magnetic fields, and accounting for unresolved density substructure (especially in the intermediate density range of $\sim10-100\ \rm{cm}^{-3}$) in order to obtain a more realistic \rmol{}. Even so, in our most promising model we still under-predict the observed median value of 0.42 by 40\%. 

\textbf{We thank the anonymous referee for their helpful and constructive comments.} We thank Chris McKee for fruitful discussions. We thank Jiayi Sun for sharing observational data sets. This work was supported by the German Science Foundation via DFG/DIP grant STE/1869-2 GE 625/17-1, 
by the National Science and Technology Council (NSTC) of Taiwan under Grant No. NSTC 113-2112-M-002-041-MY3 and National Taiwan University under Grant No. NTU-NFG-114L7444,
by the Center for Computational Astrophysics (CCA) of the Flatiron Institute, and the Mathematics and Physical Sciences (MPS) division of the Simons Foundation, USA.

\bibliography{library, new_library}
\bibliographystyle{aasjournal}

\end{document}